\documentclass[
reprint,
superscriptaddress,
aip,
jcp,
amsmath,amssymb,
floatfix,
]{revtex4-1}
\usepackage{dcolumn}
\usepackage{bm}
\usepackage{graphicx} 

\usepackage{float}
\usepackage{multirow}
\usepackage{subfig} 

\usepackage[nottoc,notlof,notlot]{tocbibind} 

\begin{document}
\title{Universal Programmable Quantum Circuit Schemes to Emulate an Operator}
\author{Anmer~Daskin}\affiliation{Department of Computer Science, Purdue University, West Lafayette, IN, 47907 USA}
\author{Ananth~Grama}\affiliation{Department of Computer Science, Purdue University, West Lafayette, IN, 47907 USA}
\author{Giorgos~Kollias}\affiliation{Department of Computer Science, Purdue University, West Lafayette, IN, 47907 USA}
\author{Sabre~Kais}\email[Corresponding author. Email: ]{kais@purdue.edu}
\affiliation{Department of Chemistry, Department of Physics and Birck Nanotechnology Center, Purdue University, West Lafayette, IN 47907, USA}
\affiliation{Qatar Environment and Energy Research Institute, Qatar Foundation, Doha, Qatar} 
\begin{abstract}
Unlike fixed designs, programmable circuit designs support an infinite number of operators. The functionality of a programmable circuit can  be altered by simply changing the angle values of the rotation gates in the circuit.
Here, we present a new quantum circuit design technique resulting in two general programmable circuit schemes.
The  circuit schemes can be used to simulate any given  operator by setting the angle values in the circuit. This provides a fixed circuit design whose angles are determined from the elements of the given matrix-which can be non-unitary-in an efficient way. 
We also give both the classical and  quantum complexity analysis for these circuits and show that the circuits require a few classical computations. They have  almost the same quantum complexities as non-general circuits. Since the presented circuit designs are independent from the matrix decomposition techniques and the global optimization processes used to find quantum circuits for a given operator, high accuracy simulations can be done for the unitary propagators of molecular Hamiltonians on quantum computers. As an example, we show how to build the circuit design for the hydrogen molecule.
\end{abstract}

\maketitle
\section{Introduction}
The classical logical devices can be broadly categorized as fixed and programmable devices. As we understand from their names, the circuits in a fixed logic can only support one function which is determined at the time of manufacture. 
This cannot be changed at a later day. On the other hand,  programmable devices such as PLDs and FPGAs are able to support an  infinite number of functionalities since they can be reconfigured outside of the
manufacturing environment. With this feature designers and programmers can run and simulate their test designs and algorithms. \cite{bobda2007introduction} 

Quantum computing has become a huge new interdisciplinary area  by providing different approaches and protocols to various  subfields including: 
communication, encryption, global binary optimization (see adiabatic quantum computing \cite{Farhi}), linear algebra, and so on \cite{Chuang,Kaye, Williams}; however,  programmable quantum  circuits and chip designs  like those in classical computers have remained  an open issue. 

In the circuit model of quantum computing, unitary matrix operators  represent  the algorithms or some part of the computations\cite{Daskin2}.  
Hence,  one of the fundamental issues is to have a general purpose quantum circuit or a quantum chip that can realize 
different types of  algorithms  in a fast and an efficient way. The possibility of designing universal quantum gate arrays  as a 
general purpose quantum computer has been discussed in ref.\cite{Nielsen}. It is shown that a gate array can be 
programmed  to evaluate the expectation value of a given operator\cite{Juan}. For the realization of a quantum gate, a cell 
structured quantum circuit design based on the activation and the deactivation of the gates on different qubits is proposed: 
It is shown that a combination of such cells can be used to realize a given quantum gate sequence\cite{Desousa}.   
Moreover,  different schemes of general programmable  universal quantum circuits are shown for two \cite{Vidal,Zhang} 
and three qubits\cite{mottonen,Vatan,Rui}  found by applying different decomposition schemes to a given unitary operator. 
Based on the general two-qubit  circuit design, a two-qubit quantum processor is experimentally realized\cite{Hanneke}.  
However, the realization of a general quantum processor and a full-scale quantum computer is still an obstacle  which requires new theoretical and experimental improvements\cite{Lanyon}.   

It is known that the realization of  quantum logical operations can be simplified by using the higher dimensional Hilbert 
spaces\cite{Lanyon,Bary}. In this paper,  using ancilla qubits, we describe  a new  circuit design approach which produces 
two programmable quantum circuit designs. These can be further improved to design general large-scale quantum
 chips and programmable quantum gate arrays. The circuits also support simulation of non-unitary matrices. 
We also show the complexity analysis for the circuits: in terms of quantum complexity, they have about the same  complexity as  non-programmable designs  
which are generated by using  matrix decompositions in numerical 
linear algebra such as QR decomposition\cite{Golub}, the quantum Shannon decomposition, the cosine-sine decomposition 
and some others \cite{Peter,vivek} (see  ref. \cite{vivek} for the comparison and the complexities of these methods).
In terms of classical complexity, since  angles for our programmable circuits can be determined simply individual matrix elements, the classical complexity is much simpler than the decomposition methods.

This paper is organized as follows:   After giving the general simulation idea, the details of two circuit designs implementing 
this idea are presented. Then the complexity of the circuits are analyzed in terms of classical and quantum complexities. Finally, we discuss the circuit designs and possible future directions. In the appendix, more computational details related to matrices are presented.

\section{The General Simulation Idea}
\begin{figure}[h]
\centering
\includegraphics[width=2.5in]{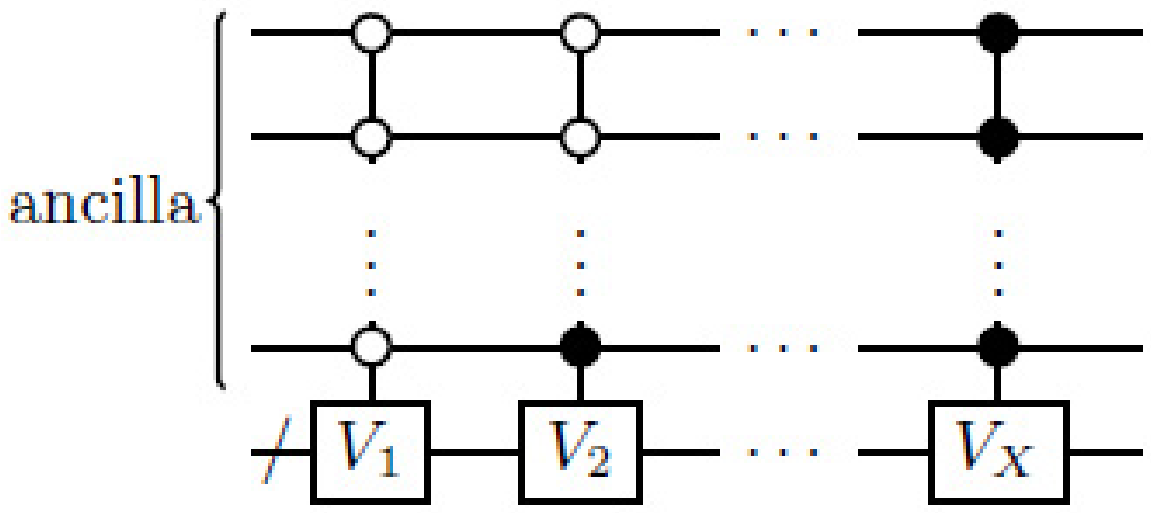}
  \caption{The number of qubits on the ancilla determines the number of $V_i$s and hence the size of $V$ in Eq.(\ref{eqbm}). }
\label{fig1}
\end{figure}

For a given real unitary $U^{N\times N}$ with $N=2^n$, and n is the number of qubits, the relationship between the input  
$|\psi\rangle=\alpha_1|0\dots0\rangle+\dots+\alpha_N|1\dots1\rangle$
 and  the output   $|\varphi\rangle$  is defined as
$|\varphi\rangle=U|\psi\rangle$ generating $N$ states:
\begin{equation}
U|\psi\rangle=\left(\begin{matrix}u_{11}&\dots&u_{1N}\\
\vdots& &\vdots\\
u_{N1}&\dots&u_{NN}\end{matrix}\right)\left(\begin{matrix}\alpha_{1}\\ \vdots\\\alpha_{N}\end{matrix}\right)=\left(\begin{matrix}\beta_{1}\\ \vdots\\\beta_{N}\end{matrix}\right).
\label{inout}
\end{equation}
Any system of higher dimension ( ancilla qubits are added to the original system) can also be used  to generate this output  
on  $N$  chosen  states with some normalization. Our  goal is to create a matrix $V$ (shown in Eq.\ref{eqbm}) which  
represents the system with the ancilla. We  then modify the initial  input $|\mathbf{0}\rangle|\psi\rangle$ to this extended 
system $V$ (the initial state of ancilla is taken as $|\mathbf{0}\rangle$ ) by using quantum operations  such that the 
application of $V$ to this modified input $|\tilde{\psi}\rangle$  includes the  output given in Eq.(\ref{inout}) with a 
normalization constant $\kappa$:   
\begin{equation}
V|\tilde{\psi}\rangle=\left(\begin{matrix}V_1&&&\\
&V_2&&\\
&&\ddots&\\
&&&V_X\end{matrix}\right)|\tilde{\psi}\rangle=\left(\begin{matrix}\kappa\beta_{1}\\ \vdots\\
\kappa\beta_2\\\vdots\\ \kappa\beta_{N}\\\vdots\end{matrix}\right),
\label{eqbm}
\end{equation}
where each $V_i$ has some distinct rows of $U$ as their leading rows. Adding a sufficient number of ancilla qubits to 
control each $V_i$ uniformly (as shown in Fig.\ref{fig1}) permits us to produce the circuit equivalent of matrix $V$  in the 
above equation.  If we assume that the first row of $V_i$ is (or includes) the $i$th row of $U$, then we need to use $ (X=N )$ such $V_i$ blocks  as shown in Eq.(\ref{eqbm}). 

The quantum operations  
to construct the matrix $V$ and the 
operations to modify the input $|\mathbf{0}\rangle\otimes|\psi\rangle$ form the circuit that simulates the given operator.
That means,  steps to form rows of $U$ in $V$ and also to  transform $|\mathbf{0}\rangle|{\psi}\rangle$ to $|
\tilde{\psi}\rangle$  generate the general circuit design for the simulation of $U$. One way to formulate these steps and to build $V_i$ matrices and the input $| \tilde{\psi}\rangle$ is as follows:
First,  the system is extended  by adding  auxiliary qubits. These ancilla qubits uniformly control different block quantum operations, $V_i$s, on the main $n$ qubits (in this paper,  $n$ or $(n+1)$ 
number of auxiliary qubits are used).
 After the formation of all elements of U which we call the \textit{Formation} step, the same row elements of $U$ are brought to the first row of each $V_i$ which we call the \textit{Combination} step.  
 The input is modified ($|\mathbf{0}\rangle|\psi\rangle\to|\tilde{\psi}\rangle)$  by a small circuit such that  $V|\tilde{\psi}\rangle$ produces an output which  includes the normalized $N$ states expected from the operation $U|{\psi}\rangle$. We call this step the \textit{Input modification} step. 
The measurement results for these  $N$ states exactly simulate $U|\psi\rangle$. The circuit design  to be found with these 
steps can be drawn as a block circuit diagram (as shown in Fig.\ref{figbc}). This approach provides a new way to find  
circuit designs. Hence,  we will describe two different programmable circuit schemes based on the block circuit in Fig.\ref{figbc}.

\section{Generation of Programmable Circuits}
\subsection{The First Circuit Design}
In this design, first we create all elements of $U$ at the diagonal positions of $V$ by using one rotation gate for each element of $U$, \textit{Formation} step. In the \textit{Combination} step,  the elements on  each $i$th row of $U$ are collected in the first row of each $V_i$.

\begin{figure*}
\centering
\includegraphics[width=3.5in]{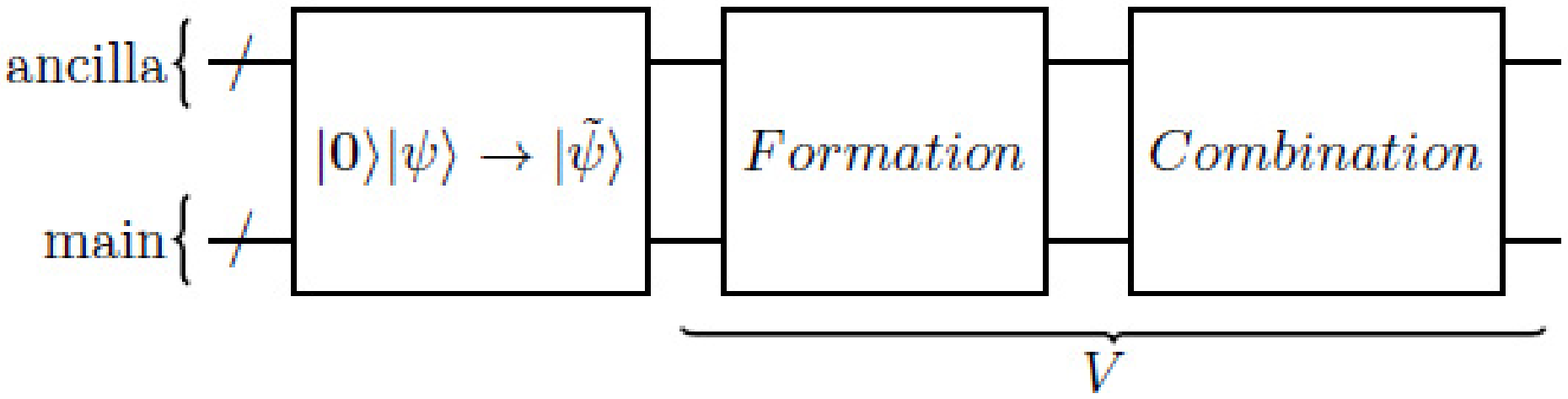}
  \caption{Block circuit diagram to simulate $U$ by modifying the input $|\mathbf{0}\rangle|\psi\rangle$ to $|\tilde{\psi}\rangle$ and constructing $V$ in two steps: the formation of the elements of $U$ in $V$ and bringing the same row elements in $U$ to the first rows of $V_i$s in $V$, combination. The necessary gates to form $V$ and to also transform $|\mathbf{0}\rangle|\psi\rangle$ to $|\tilde{\psi}\rangle$ will generate the circuit.}
\label{figbc}
\end{figure*}

\textbf{\textit{Formation} Step:} In this step,  the elements of $U$ are tiled across the diagonal of a new higher-dimensional matrix $V_f$. This is a block diagonal matrix with $2 \times 2$ blocks across the diagonal. For each element of $U$, one rotation gate is used. The angular value for the gate is determined to form an element of $U$ as its cosine value. Controlling such gates in a uniform binary coded fashion produces the matrix which has all elements of $U$ on its diagonal:
\begin{equation}
V_f=\left(
\begin{matrix}
R_1&\\
&\ddots&\\
&&R_{N^2}	
\end{matrix}
\right)_{2N^2\times2N^{2}}, R_j=\left(\begin{matrix}c_j&s_j\\-s_j&c_j\end{matrix}\right),
\label{Eqmat1}
\end{equation}
where $c_j=cos(\arccos(u_j))$ generating the $j_{th}$ element of $U$.  We use  $(n+1)$ number of ancilla qubits to uniformly control  each $R_j$, $1\le j\le N^2$.

\textbf{\textit{Combination} Step:}   To bring the same row elements of $U$ to the first rows of the $V_i$s, we need a quantum operation $V_c$ which will produce the matrix $V=V_cV_f$ represented as:
\begin{widetext}
\begin{equation}
\left(\begin{matrix}
K\\
&K\\
&&\ddots\\
&&&K\\
\end{matrix}\right)
\left(\begin{matrix}
         \begin{matrix}c_1&s_1\\-s_1&c_1\end{matrix}&\\
          &\ddots&\\
          &&\begin{matrix}c_{N^2}&s_{N^2}\\-s_{N^2}&c_{N^2}\end{matrix}
          \end{matrix}
\right)
=
\left(\begin{matrix}
\begin{matrix}
ku_{11}&\cdot&ku_{12}&\dots & ku_{1N}\\
 \vdots&\vdots&\vdots&&\vdots
\end{matrix}&  &
\\ 
& \ddots  &
\\
 &  &\begin{matrix}
ku_{N1}&\cdot&ku_{N2}&\dots & ku_{NN}\\
 \vdots&\vdots&\vdots&&\vdots
\end{matrix}
\end{matrix}\right),
\label{Eqmat2}
\end{equation}
\end{widetext}

where $K$ should have a form similar to the following matrix:

\begin{equation}
K=\left(\begin{matrix}
k&0&k&\dots&0& k &0\\
0&k&0&\dots&k&0&k\\
k&0&k&\dots&0&k&0\\
\vdots&\vdots&\vdots&\vdots&\vdots&\vdots&\vdots\\
0&k&0&\dots&k&0&k\\
k&0&k&\dots&0&k&0\\
0&k&0&\dots&k&0&k\\
\end{matrix}\right)_{2N\times2N}.
\end{equation}
 For  a system with  (n+1) qubits, the single Hadamard gates on the first $n$ qubits generate the above matrix 
with $k=\pm1/\sqrt{2^n}$. Hence, $V_c$ is the matrix form of this operation in the system with $(2n+1)$ qubits where we apply the Hadamard 
gates to the $(n+1)$st, $n$th, ..., $3$rd, and $2$nd qubits from the bottom in the circuit.

\textbf{\textit{Input modification} Step:} In the final matrix in Eq.\ref{Eqmat2}, since the corresponding state for the rows 
which posses the elements of $U$ with the normalization factor $k$ are to be assigned as   $N$ chosen states simulating  
$U$, we should modify the input in such a way that the elements represented as 
\lq\lq{}$\cdot$\rq\rq{}s between $ku_{ij}$ and $ku_{i(j+1)}$ are disregarded. That means the initial input should be 
transformed into $|\tilde{\psi}\rangle$ by a prior operation to the final matrix $V$ so that the corresponding elements in 
the input to  \lq\lq{}$*$\rq\rq{} elements are set to  zero:

\begin{equation}
\begin{split}
 &|\mathbf{0}\rangle|\psi\rangle\rightarrow|\tilde{\psi}\rangle=\\ &[\alpha_1\ \alpha_2\ \dots \alpha_N\ 0 \dots 0]^T \rightarrow
\\  &[\kappa\alpha_1\ 0\ \kappa\alpha_2\dots 0\ \kappa\alpha_N\dots\ \kappa\alpha_1\ 0\ \kappa\alpha_2\ \dots\ 0\ \kappa\alpha_N]^T,
\end{split}\end{equation} 
where $\kappa$ is a normalization constant.  It is easy to see that this modification can be succeeded by simple Hadamard gates on the first $n$ qubits, and sequential swap operations between the (n+1)st  and the remaining  $n$  qubits. 

\begin{figure*}[t]
  \centering
  \includegraphics[scale=0.8]{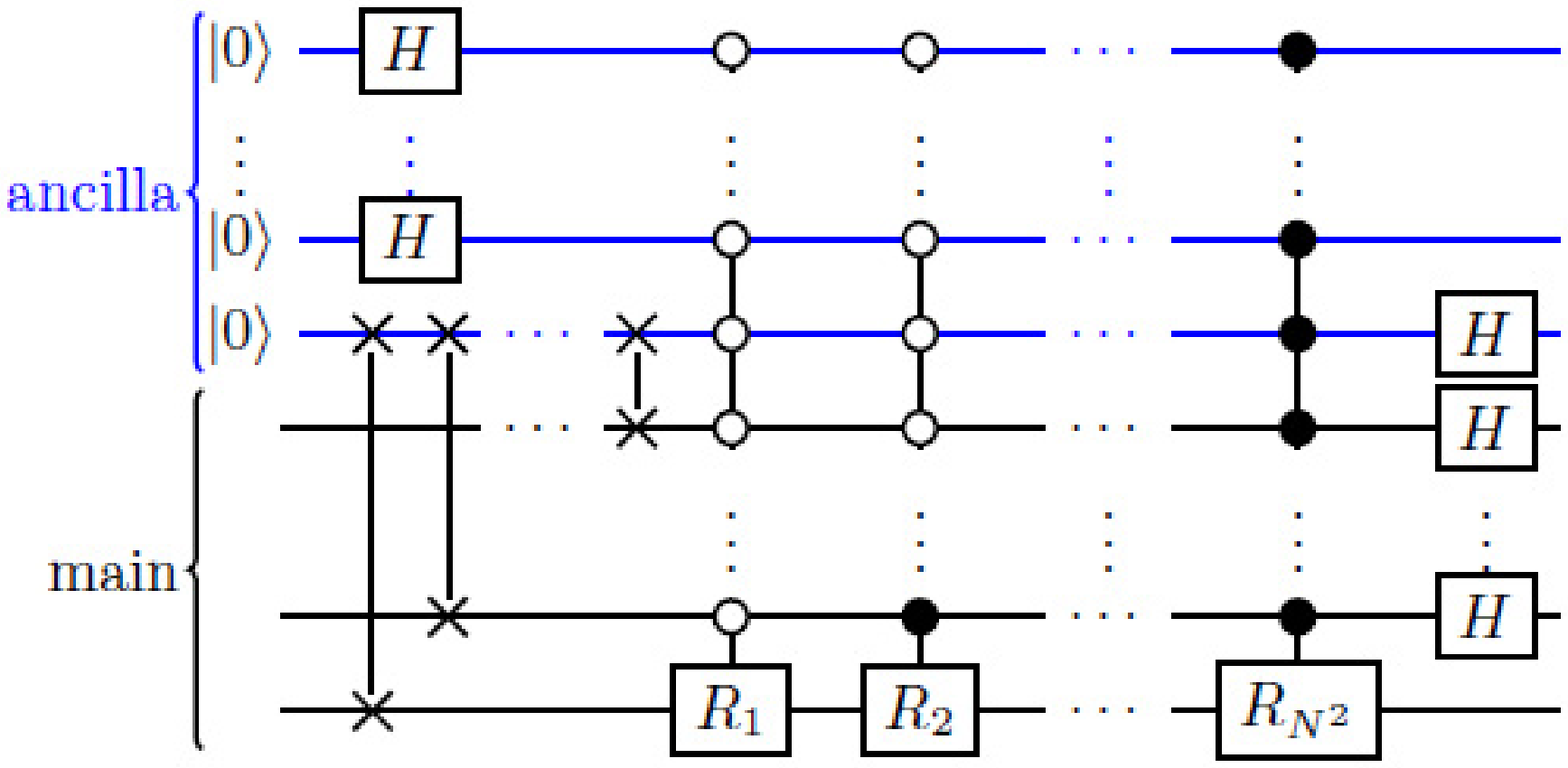}
  \caption{
  The first  circuit design for a given general matrix: the initial Hadamards and the SWAPs are to modify the input, and the last Hadamards carry the elements to the  first rows of $V_i$s (combination step). The uniformly controlled quantum gates in the middle form all elements of $U$ on the diagonal of $V$ (formation step).
  }
\label{circuit1full}
\end{figure*}

The equivalent circuit simulating any $U$ is drawn in Fig.\ref{circuit1full} for $n$ qubit system by adding $n+1$ ancilla qubits and replacing the block circuits in Fig.\ref{figbc} with the explicit circuits found above. 

At the end of this circuit, which can be decomposed into one- and two-qubit gates by using the decomposition technique discussed in Sec.\ref{sec:complexity}, the following set of $N$ states exactly simulates the given unitary $U$ after normalization:
\begin{equation}
\begin{split}
|0\dots000-0\dots0\rangle,\\
|0\dots010-0\dots0\rangle,\\
\vdots\\
|1\dots110-0\dots0\rangle,
\end{split}\end{equation} where the dashes are used to separate the main and the ancillary. 

In Appendix \ref{ap}, we give an example of the explicit matrices used for each step of the algorithm.

\subsection{The Second Circuit Design}
In the first circuit design, the elements of $U$ are initially formed on the diagonal of $V$ by using uniformly controlled rotation gates. Here,  we take a group of elements from a row of $U$ and create them as the leading row of  small block matrices by preserving the ratios between the elements.
Using a rotation gate for each two of these initial small blocks,  we create  larger block matrices which will have more  elements of $U$ in their first rows. 
This combination of steps  is iteratively done until  the final $V_i$s with leading rows having the rows of $U$  as in Eq.(\ref{eqbm}) are constructed. Since the final blocks, $V_i$s, are  $N\times N$, the matrix $V$ is $N^2\times N^2$; therefore $n$  qubits are needed for the ancilla. The input modification step follows the same idea as described for the first design.

\begin{figure*}
  \centering
  \includegraphics[width=5.5in]{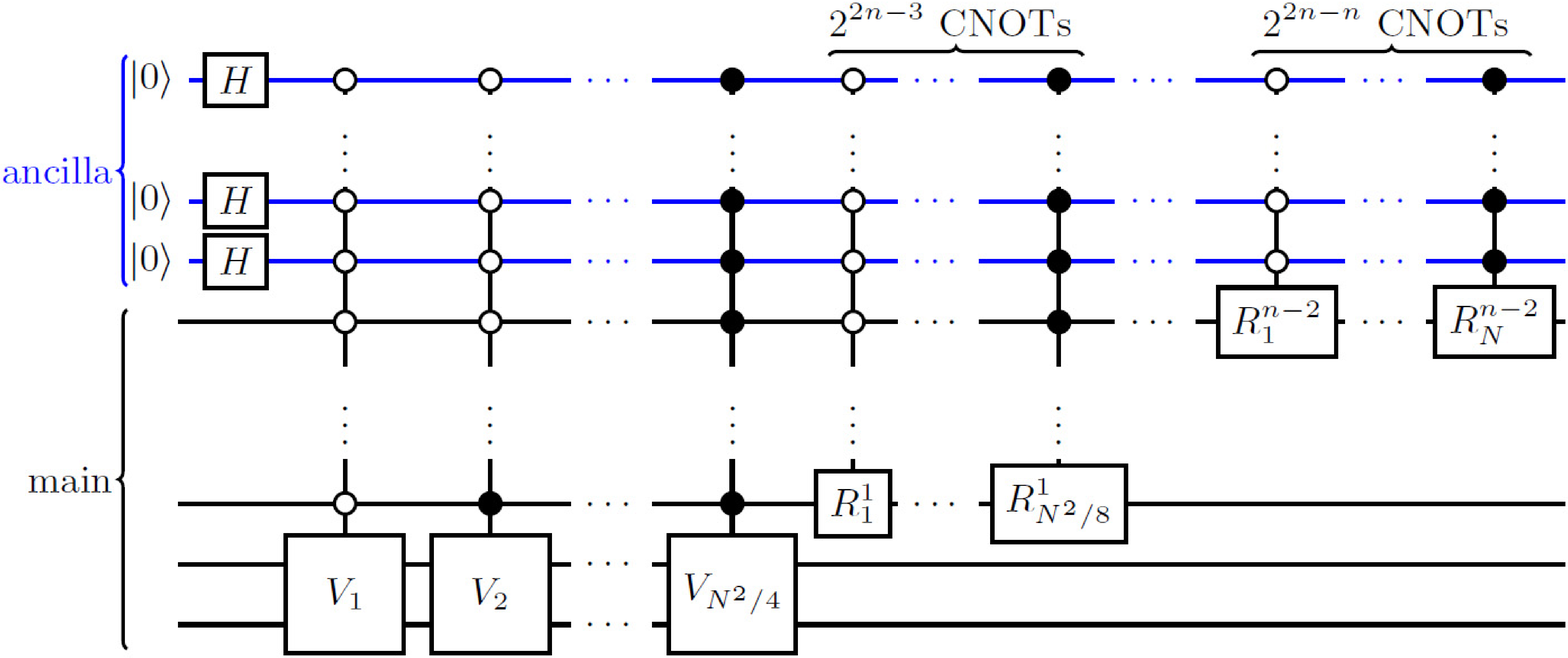}
  \caption{
The second circuit with  4 by 4 initial blocks: The differently controlled quantum gates in the networks, after the $V_i$ blocks,  combine small blocks and build the $N$ by $N$ blocks at the end.  The initial Hadamards are for the modification of the input. The $V_i$ blocks are for the formation step.
  }
\label{circuit2full}
\end{figure*}

\textbf{Formation Step:}
As stated above, instead of forming  matrix elements at the diagonal positions by using a rotation gate for each element of $U$, 
 a group of elements is created in the first row of each block with the same ratio as those elements  in the original matrix. For instance, if the initial blocks are of dimension 2 by 2, the first row implements  two elements, $u_{ij}$ and $u_{ik}$, of $U$. Thus, the ratio between the first element and the second element of a 2 by 2 block matrix is the same as $u_{ij}/u_{ik}$ (since the block is 2 by 2, the elements of the block matrix are the cosine and sine values of an angle $\theta_x$ which provides the equality $\cos(\theta_x)/\sin(\theta_x)=u_{ij}/u_{ik})$. 
 In our circuit designs, we will assume $k=j+1$, and so the first row elements of each block implement the ratios of the elements  in the same order as the original matrix. Therefore, if the first blocks are of dimension $d\times d$;  the total number of initial blocks will be $N/d$ since each block implements $d$ number of elements.    
The following matrix represents the formation step for 2 by 2 initial blocks:
\begin{widetext}
\begin{equation}
V_f=\left(\begin{matrix}
\begin{matrix}
k^1_1u_{11}& k^1_1u_{12}\\
- k^1_1u_{12}&k^1_1u_{11} \\ 
\end{matrix}
\\  &\ddots\\
&&\begin{matrix}
k^1_{\frac{N}{2}}u_{1N-1}& k^1_{\frac{N}{2}}u_{1N}\\
-k^1_{\frac{N}{2}}u_{1N}&k^1_{\frac{N}{2}}u_{1N-1}\\ 
\end{matrix}\\
&&&\ddots\\
&&&& \begin{matrix}
k^N_{\frac{N}{2}}u_{NN-1}& k^N_{\frac{N}{2}}u_{NN}\\
- k^N_{\frac{N}{2}}u_{NN}&k^N_{\frac{N}{2}}u_{NN-1} \\ 
\end{matrix}
\end{matrix}\right),
 \label{matrix3}
\end{equation}\end{widetext} where $k^i_j$s are the normalization constants, and $u_{ij}$s are the elements of $U$.
 The $V_i$ block operations in Fig.\ref{circuit2full} produce a matrix $V_f$ with 4 by 4 block matrices on its diagonal.
 
\textbf{Combination Step:} After the formation with ratios,   blocks are combined  using one rotation gate for each pair of two blocks so as to form new  larger blocks with new 
normalization constants that preserve the original ratios of the elements. Each of these new blocks  has  twice as many elements as the former blocks.  As an example,  we will combine two 
4 by 4  matrices located on the diagonal of the matrix $V_8$:
\begin{equation}
V_8=\left(\begin{matrix}
\begin{matrix}
k_1u_{1}&\dots & k_1u_{4}\\
\cdot&\dots&\cdot\\
\cdot&\dots&\cdot\\
\cdot&\dots&\cdot
\end{matrix}&
\\  &
\begin{matrix}
k_2u_{5}&\dots & k_2u_{8}\\
\cdot&\dots&\cdot\\
\cdot&\dots&\cdot\\
\cdot&\dots&\cdot 
\end{matrix}
\end{matrix}\right),
 \label{matrix3}
\end{equation}
where and $k_1$ and $k_2$ are the normalization factors. The following matrix, $V_{c_8}$, can be used  as a combination matrix  to generate  an 8 by 8 larger  block from the above pair of two 4 by 4 blocks:
\begin{equation}
V_{c_8}=\left(\begin{matrix}
c_x&0 &0&0&s_x&0&0&0\\
0&c_x &0&0&0&s_x&0&0\\
0&0&c_x&0 &0&0&s_x&0\\
0&0&0&c_x& 0&0&0&s_x\\
-s_x&0&0&0&c_x&0&0&0\\
0&-s_x&0&0&0&c_x&0&0\\
0&0&-s_x&0&0&0&c_x&0\\
0&0&0&-s_x&0&0&0&c_x\\
\end{matrix}\right)
\end{equation}
where $c_x=cos(\theta_x)$,  $s_x=\sin(\theta_x)$, and $\theta_x$ is an angle to achieve the required ratio. The matrix multiplication $V_{c_8}V_8$ produces a matrix with the leading row $[k_xu_{1}\ \dots \ k_xu_{8}]$, where $k_x=sin_x\times k_2$ and $k_x=cos_x\times k_1$. 

It is easy to see that the matrix $V_{c_8}$ can be written as $R(2\theta_x)\otimes I\otimes I$. Hence, any such general combination matrix can be written as $R\otimes I^D$ where $D$ is the size of the blocks to be combined by using $V_c$; and R is a general one qubit rotation gate. This means that for the blocks operating on $c$ qubits, if we apply a rotation gate to the $(c+1)$st qubit, it will be equivalent in matrix form to the operation $V_cV_{2^{c+1}}$. Hence, putting single rotation gates on $(c+1)$st, $(c+2)$nd, ..., $n$th qubits generates an $N$ by $N$ matrix. Furthermore, by controlling each $V_c$  operation (or equivalently single rotation gates, $R$s) uniformly by the upper qubits in the circuit (see the uniformly controlled rotation gates located after the $V_i$ block operations in Fig.\ref{circuit2full}), we can generate $N$ such separate blocks and  the following final matrix:

\begin{equation}
V=\left(\begin{matrix}
\begin{matrix}
u_{11}&\dots & u_{1N}\\
 \vdots&\vdots&\vdots 
\end{matrix}& & 
\\
&  \ddots   &
\\
 & & \begin{matrix}
u_{N1}&\dots & u_{NN}\\
 \vdots&\vdots&\vdots 
\end{matrix}
\end{matrix}\right).
\label{matrix1}
\end{equation}
Since the resulting rows in each block are unit vectors and have the same ratio as the row elements of $U$, they are equal to the corresponding rows of $U$. (The final normalization constants become equal to 1.)

For the general case, if the initial  blocks are operating on the last $c$  qubits,
  we need to use $N/2^c$ uniformly controlled rotation gates on each main qubit (excluding the last $c$ qubits) 
in order to recursively combine small blocks.  At the end, we   have $N\times N$ blocks whose leading rows are the rows of $U$ as shown in Eq.( \ref{matrix1} ).

\textbf{Input Modification ($|\bold{0}\rangle|\psi\rangle\rightarrow|\tilde{\psi}\rangle$) :} Modification of the input $[\alpha_1\ \alpha_2\ \dots \alpha_N\ 0 \dots 0]^T$ as $[\kappa\alpha_1\dots \kappa\alpha_N|\kappa\alpha_1\dots \kappa\alpha_N|\dots|\kappa\alpha_1\dots \kappa\alpha_N]^T$ with  the normalization constant $\kappa$ allows us to simulate $U$ by using $V$ in Eq.(\ref{matrix1}) on the chosen $N$ states: 
\begin{equation}
\begin{split}
|0\dots000-0\dots0\rangle,\\
\vdots \\
|1\dots111-0\dots0\rangle.
\end{split}
\end{equation} 
This input with $\kappa=1/\sqrt{2^N}$ can be  produced by applying the Hadamard gates to all ancilla qubits at the beginning of the circuit. 

Consequently,  the general circuit design shown in Fig.\ref{circuit2full} is obtained which is able to simulate any real unitary matrix. For more explicit matrix forms and illustrative details, please refer to Appendix \ref{ap} and Appendix \ref{ap2}.

\section{Complexity Analysis of The Circuits}
\label{sec:complexity}
\begin{figure*}
  \subfloat[]{\centering
  \includegraphics[scale=0.3]{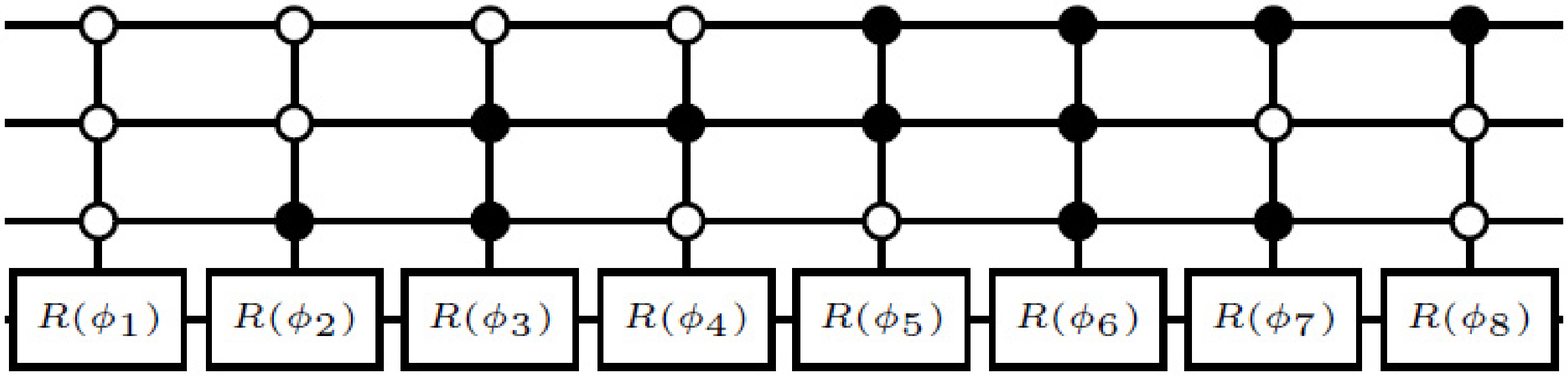}
\label{circuit3}
}

\subfloat[]{ \centering
\includegraphics[scale=0.3]{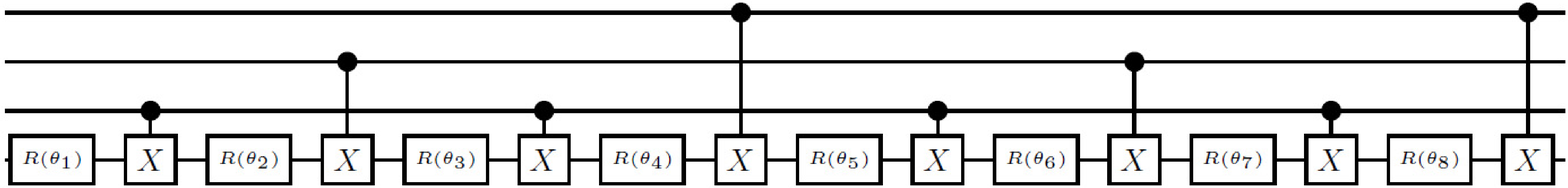}
\label{circuit4}}
  \caption{
  (a) A gray-coded multi-control network. (b) The decomposition of the gray-coded network in (a) into CNOT and single quantum gates.
  }
\label{circuit3full}
\end{figure*}

In the cases of classical and quantum complexities of the circuits explained above, it is easy to see that they  depend on  mostly the costs of uniformly controlled networks such as the one in Fig.\ref{circuit3}. Such a network controlled by $k$ qubits can be decomposed in terms of $2^k$ CNOT gates and $2^k$ single rotation gates\cite{mottonen}.  For instance, the circuit  as illustrated for $k=2$ in Fig.\ref{circuit3} can be decomposed  as in Fig.\ref{circuit4}.
The angle values in the decomposed circuit are found to be the solution of the system of the  linear equation $M^{k}\boldsymbol\theta=\boldsymbol\phi$:
\begin{equation}
M^{k}\left(\begin{array}{c}
\theta_1\\
\theta_2\\
\vdots\\
\theta_{2^{k}}
\end{array}\right)=\left(\begin{array}{c}
\phi_1\\
\phi_2\\
\vdots\\
\phi_{2^{k}}
\end{array}\right),
\end{equation} where  k is the number of control qubits in the network, and the entries of $M$ are defined as:
\begin{equation} 
M_{ij}=(-1)^{b_{i-1}.g_{j-1}},
\end{equation}
 in which the power term is found  by taking the dot product of the standard binary code of the index $i-1$, $b_{i-1}$,  and the binary representation of $j-1$th gray coded integer, $g_{j-1}$. Since  $M^k$ is a column permuted version of the Hadamard matrix, we see that $M$ is unitary. Thus,  $(M^k)^{-1}=2^{-k}(M^k)^T$, and the new angle values in the decomposed circuit are the result of the mere matrix vector multiplication\cite{mottonen}: 
\begin{equation}
\boldsymbol\theta=2^{-k}(M^k)^{T}\boldsymbol\phi.
\label{matvec}
\end{equation}

\subsection{The complexity of the first circuit design}
\subsubsection{The Classical Complexity}
In the first circuit diagram (see Fig.\ref{circuit1full}), since there is only one such network,  we need to multiply the $2^{2n}\times 2^{2n}$ matrix  by the vector of  dimension 
$2^{2n}$ constructed by taking the arc-cosines of  every element of $U$. Hence, the classical complexity for the first circuit is $O(2^{4n})$. However, since $M$ is the permuted version 
of the Hadamard matrix, by using the fast Hadamard transform \cite{Fino}, which requires $O(NlogN)$ computations for the transform of a vector by the Hadamard matrix, this can be achieved in:
\begin{equation}
O(2^{2n}log(2^{2n}))=O(2n2^{2n}).
\end{equation}

\subsubsection{The quantum Complexity}
 The quantum complexity of the circuit is the number of gates required for the decomposition of the  network,  the combination of the blocks and the input modification: $2^{2n}$ CNOT, $2^{2n}$ single rotation, $2n$ Hadamard, and $n$ SWAP gates.

\subsection{The complexity of the second circuit}
\begin{figure*}
 \subfloat[]{ 
 \centering \includegraphics[scale=0.3]{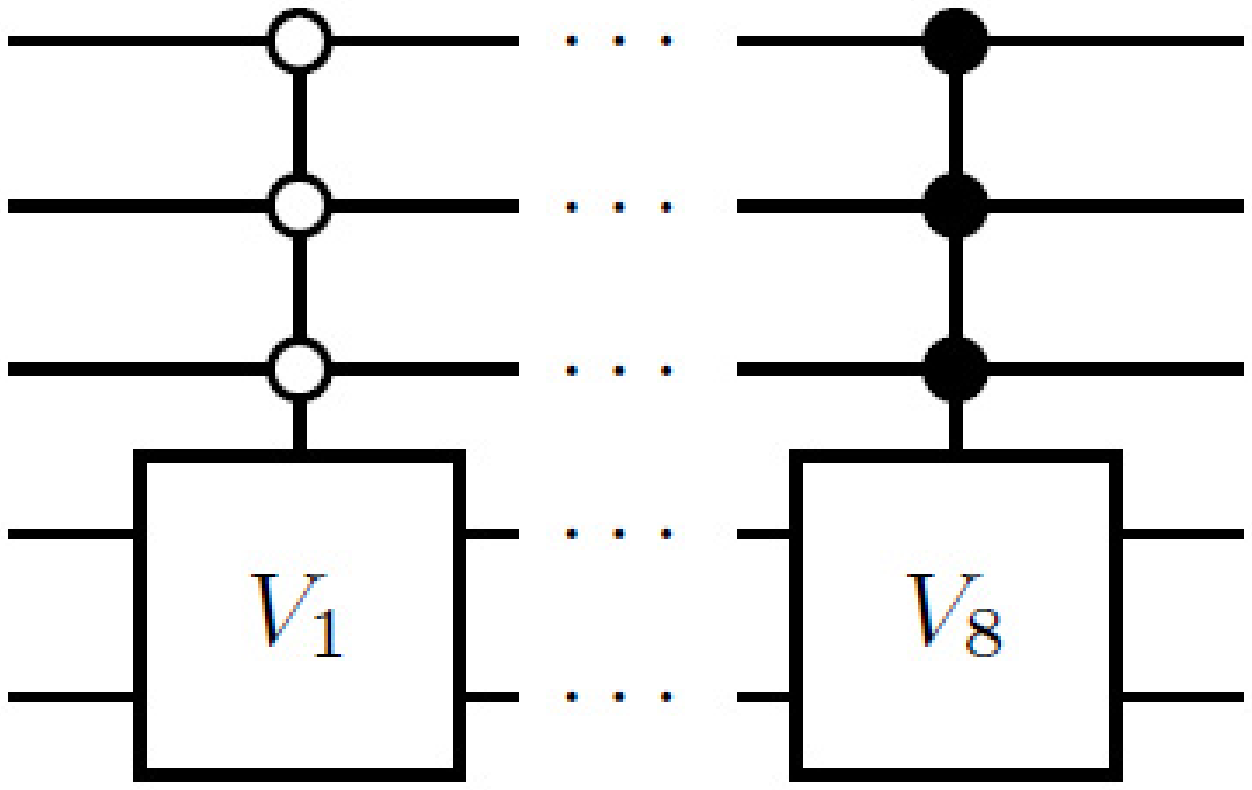}
 \label{block}
}

\subfloat[]{ \centering \includegraphics[scale=0.3]{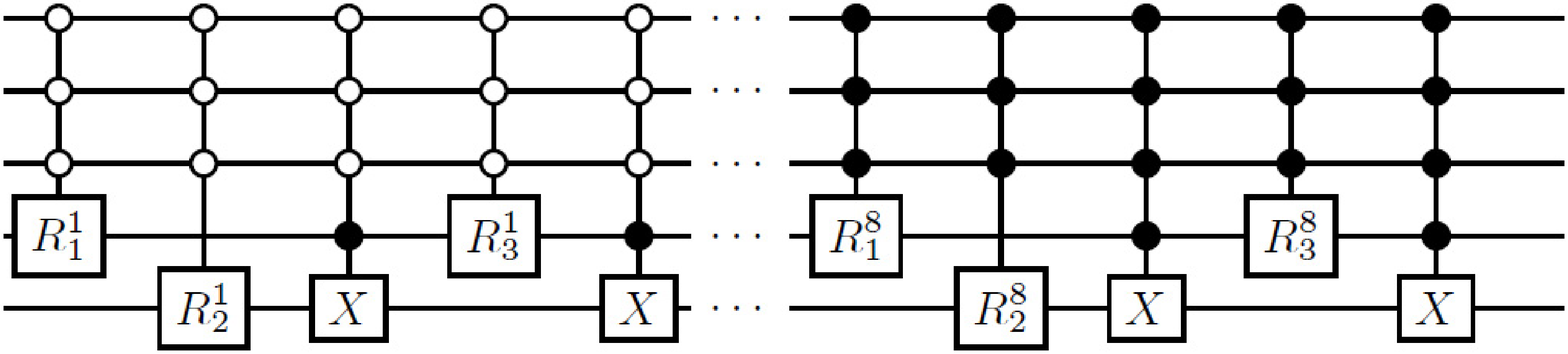}
\label{opened}
  }

\subfloat[]{ \centering \includegraphics[scale=0.3]{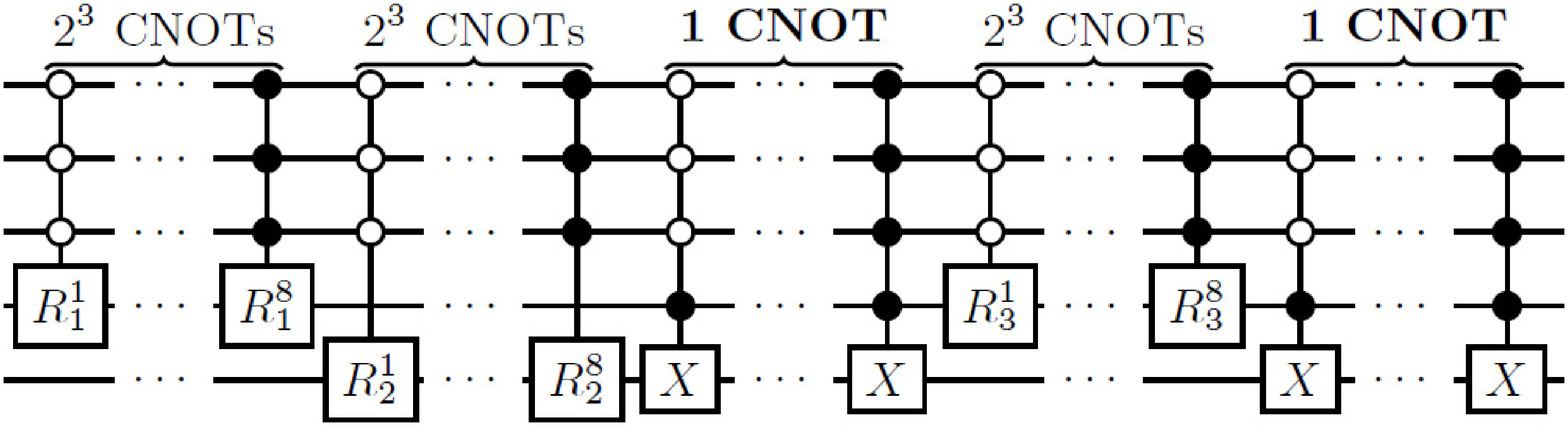}
\label{combined}
  }
  \caption{
The circuit in (a) with  4 by 4 initial blocks can be represented as in (b) by using the circuit given in Fig.(\ref{schmidt}). Without changing the order of the gates having the same control state, the gates can be moved to form uniformly controlled networks as in (c): If a gate has the same angle value for all control states such as the control $X$ gates in the circuit, they are equal to a single gate (in the case of $X$ gates in the circuit, only one CNOT is required).
  }
\label{blockopened}
\end{figure*}
 The classical  and the quantum complexities  for the second circuit are determined by the number of networks which are formed by  putting the quantum gates in blocks controlled uniformly together as shown in Fig.\ref{blockopened} and by  the combination steps. Since the quantum gates in different blocks with the same angles   operate for every case of the control qubits, putting them together do not produce  networks. Instead,  they need to be applied only once such as the controlled $X$ gates shown in Fig.\ref{combined}.  Hence, if  the initial blocks of $2^c$ by $2^c$ (operating on $c$ qubits) include $m$ different quantum gates (the type of the gates are the same, but each requires different angles in different blocks such as $R^1_1$ and $R^8_1$ in Fig.\ref{blockopened}),  these blocks together produce $m$ gray coded networks controlled by $2^{2n-c}$ qubits. 

In addition, in the combination step, we use binary coded networks on each main qubit excluding the last $c$ qubits to produce $N$ by $N$ blocks. Thus, we will also have $n-c$ gray coded networks for the combination step  for which the numbers of control qubits go down by one from one combination step to another(or from one gray-coded network to another). The classical and the quantum complexity will be determined mostly by the decompositions of these $m+(n-c)$ networks.

\subsubsection{Classical Complexity}
As mentioned above, in the formation step, the combination of  decomposed block circuits together form $m$ gray coded networks for $m$ different gate as represented for two-qubit blocks in Fig.\ref{blockopened}. Hence, to find the decompositions of these networks as in Fig.\ref{circuit4} by the formula given in Eq.(\ref{matvec}), $m$ number of matrix-vector  multiplications are needed: The dimensions of the matrices are  $2^{2n-c} \times 2^{2n-c}$ and the dimensions of the vectors are $2^{2n-c}$. Using the fast Hadamard transform,  the complexity for this part is found to be $O_f=O(m(2n-c)(2^{2n-c}))$ instead of $O(m(2^{2n-c})^2)$ by the naive matrix vector multiplication. 

Furthermore, the combination step is the summation of the computations done for finding the angles of $(n-c)$ gray coded networks (remember that the number of control qubits decreases by one from one network to another).
This is equal to $O((2^{2n-c-1})^2)+O((2^{2n-c-2})^2)+\dots+O((2^{2n-c-n+c})^2)=O(2^{4n-2c}-2^{2n})$ by the naive matrix vector multiplication.
By the fast Hadamard transform, the complexity of the combination step is  as follows:
\begin{widetext}
\begin{equation}
\begin{split}
O_c=&O((2n-c-1)(2^{2n-c-1}))+O((2n-c-2)(2^{2n-c-2}))+\dots+O(n2^{n})\\
=&O(2\times (1-(2n-c)2^{2n-c-1} + (2n-c-1)2^{2n-c}))-O(2\times(1-n2^{n-1}+(n-1)2^{n}))\\
=&O((2n-c-2)2^{2n-c}-(n-2)2^n)
\end{split}
\end{equation} 
\end{widetext}
Thus, while the total complexity by the naive multiplication is \begin{equation}
\begin{split}
&O(2^{4n-2c}-2^{2n})+O(m(2^{2n-c})^2)\\
&=O((m+1)2^{4n-2c}-2^{2n}),
\end{split}
\end{equation}
 by the fast Hadamard transform, it is:
\begin{widetext}
\begin{equation}
\begin{split}
O_f+O_c=
O((m+1)(2n-c)2^{2n-c}-2^{2n-c+1}-(n-2)2^n)).
\end{split}
\end{equation}
\end{widetext}

\subsubsection{The Quantum Complexity }
In terms of the quantum complexity, the analysis follows the same structure: as mentioned, $m$ different gates in the blocks on $c$ qubits create $m$ networks controlled by $2n-c$ qubits. The decomposition of these networks requires $m2^{2n-c}$ CNOT and the same number of single gates. 

Since $n-c$ combinations ($n-c$ network) are necessary, the complexity of the combination step is the summation of $n-c$ terms: $2^{2n-c-1}+2^{2n-c-2}+\dots+2^{2n-c-n+c}=2^{2n-c}-2^n$.

Then the total CNOT complexity reads as:
\begin{equation}
2^{2n-c}-2^n+m2^{2n-c}+\Phi=(m+1)2^{2n-c}-2^n+\Phi
\end{equation}
 where $\Phi$ represents the common gates in each block that needs to be run only once.

\begin{center}
  \begin{figure}
\centering
\includegraphics[scale=0.3]{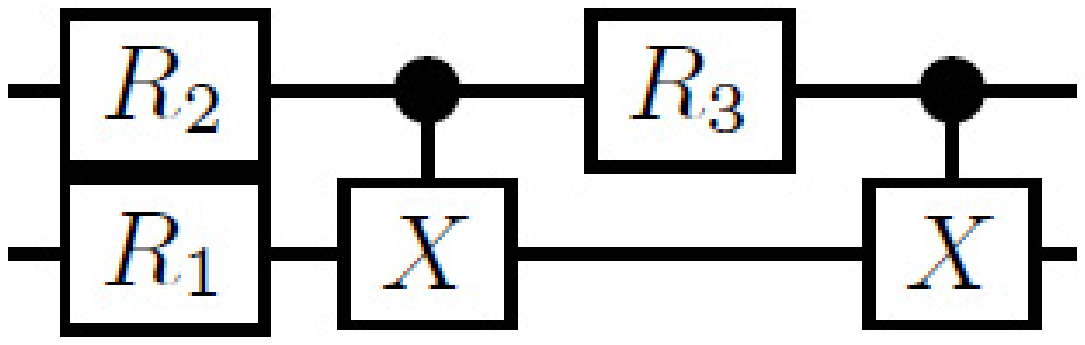}
\caption{Quantum circuit which is found by following the Schmidt decomposition and can generate any vector of dimension 4 as the first row of its matrix representation. }
\label{schmidt}
\end{figure}
\end{center}

\textbf{Example:} As an example, the complexity of a general 4 by 4 block circuit can be found as follows:
By using the Schmidt decomposition\cite{Kaye}, any 1 by 4 unit vector $\mathbf{u_x}$ can be decomposed as:
$\mathbf{u_x}=\sum_{i=1}^{2}{a_i\mathbf{v_{i}^1}\otimes \mathbf{ v_{i}^2}}.$
Since $V_1$ and $V_2$ composed of $v_i^1$ and $v_i^2$ vectors are 2 by 2 unitary matrices, these matrices (with the elements ($cos_1$ and $sin_1$ for $V_1$, and $cos_2$ and $sin_2$ for $V_2$) and  the coefficients satisfying  $|a_1|^2+|a_2|^2=1$ can be considered as the rotation gates. For the coefficients, $a_1$ and $a_2$ are the cosine and the sine values of a rotation gate ($a_1=cos_a$ and $a_2=sin_a$). The resulting decomposition becomes equal to the following:
\begin{equation}
\mathbf{u_x^T}=\left(\begin{array}{cccc}
a_1cos_1cos_2+a_2sin_1sin_2\\
-a_1cos_1sin_2+a_2sin_1cos_2\\
-a_1sin_1cos_2+a_2cos_1sin_2\\
a_1sin_1sin_2+a_2cos_1cos_2\\
\end{array}\right),
\end{equation} which requires three rotation gates in general.
The circuit given in Fig.\ref{schmidt}  forms  any $\mathbf{u_x}$ as the leading row of its 4 by 4 matrix.
 Therefore,  taking this circuit to implement the blocks in Fig.\ref{circuit2full} gives $c=\Phi=2$, and $m=3$; hence the CNOT complexity of the whole circuit in Fig.\ref{circuit2full} reads as $2^{2n}-2^n+2.$
Also note that if the blocks in the circuit shown in Fig.\ref{circuit2full} were of dimension 2 by 2, then the complexity would be $2^{2n}-2^n$.

\subsection{Comparison with the Non-Programmable Circuit Designs}
The reported non-general circuit decompositions have the CNOT complexities ranging from $O(n^32^{2n})$ to the most efficient one $\frac{3}{4}2^{2n}-\frac{3}{2}2^n$. The 
proven lower bound for the CNOT complexity is $(2^{2n-2}-3n/4-1/4)$   without using any auxiliary qubits \cite{vivek}. Even though the circuit designs given in 
this paper are general and fixed size for any operator, their complexities are  greater  by roughly a factor of 2 compared to those nonprogrammable circuits. In addition, if we can make $m$ less than or equal to $2^{c-2}$, then we can also go below the lower bound. This is likely to happen because the common quantum gates in the blocks (as  two CNOTs in 4 by 4 blocks) do not affect the upper bound of the complexity. Hence, by benefiting from this property,  the lower bound complexity may be reduced with the use of higher Hilbert spaces.

\section{Discussion and Conclusion}
\subsection{Programmable Quantum Chips}
The circuit designs given here are independent of the type of operator; hence they can be used to design general purpose quantum processors and quantum chips in which the angles 
are set  by a preprocessing unit. They can also help in the design of possible quantum gate arrays\cite{Nielsen}. 
In addition, because the circuit designs  are highly dependent on the matrix elements, for the application specific circuits aimed to implement particular types of systems,  any level of 
sparsity in the system may reduce the number of gates significantly in the general design;  
hence, more efficient  quantum chips can be built for particular uses. For instance, if  half of each of the row elements are zero in the given matrix, considering the first approach, the 
blocks at the end of the combination steps can  be made to have  the dimension $( N/2 \times N/2 )$. Hence, this will lead the circuit to require fewer combination steps (the number of qubits in ancilla is reduced by one), which lowers the both classical and CNOT complexities and makes any possible fabrication easier.

\subsection{Finding Angles}
In the case of finding the angle values on classical computers for a given unitary operator,   the process can be  parallelized conveniently to find the angles. For instance, the distribution of each row to the different cores may be one way of parallelization of the method. This can be further  improved and designed in terms of more small blocks.  And so the computation time to generate angles for both circuits can be very fast.

The combination procedure described for both circuit designing processes can be  further improved to combine circuits for different unitary operations by considering them as initial blocks. One of the individual blocks used to generate a row of the given matrix can also be used as the state preparation circuit (for instance Fig.\ref{schmidt}) for an arbitrary circuit.  Furthermore, the circuits generated by the first approach have high resemblance to the qubus quantum computer\cite{Katherine}. Similar ideas can be used to implement circuit design techniques for   this type of quantum computers as well.
 
\subsection{Complex Cases}
 It is  important to note that in this  paper, even though  real matrices are considered, it is straightforward to implement any complex case as well by considering each rotation gate as also being able to produce any complex element of a unitary matrix in the first circuit design. This may require more than one simple rotation gate, but it shall not increase the upper bound of the quantum complexity. However, the modification for the second circuit may not be as simple as for the first one: this may require additional gates during the combination and formation steps.
 
\subsection{Simulation of Molecular Hamiltonians}
The exponential growth of computational cost with the number of atoms is a huge computational challenge for the exact quantum chemistry calculations. Even for a simple molecule like methanol,
using only the 6-31G**  basis for the valence electrons,
there are 50 orbitals. The 18 valence electrons can be distributed in these orbitals in any way that satisfies the Pauli exclusion principle. This leads to about $10^{17}$ possible configurations making an exact or Full Configuration Interaction (FCI) calculation almost impossible on classical computers\cite{Daskin2}.
However, it has been shown that a quantum computer can be used to estimate the ground and
excited state energies of  molecules efficiently\cite{Daskin2,Abrams,Alan-Science,Veis, Hefeng,Alan2,Papageorgiou,Lanyon2,Kassal,Veis2,Kassal2}. 
For the simulation of a quantum system, it is necessary to find an
equivalent  quantum circuit  to the unitary propagator of the Hamiltonian representing that system. 
The molecular electronic Hamiltonian, in the Born-Oppenheimer approximation, is described in the second quantization form as \cite{Daskin2,Lanyon,Alan2}:
\begin{equation}
\mathcal{H}=\sum_{pq}{h_{pq}a_{p}^{\dagger}a_q}+\frac{1}{2}\sum_{pqrs}{h_{pqrs}a_{p}^\dagger a_{q}^\dagger a_{s}a_{r}},
\label{h2hamiltonian}
\end{equation}
where the matrix elements $h_{pq}$ and $h_{pqrs}$ are the set of one- and
two-electron integrals, and $a_j$ and $a_{j}^\dagger$ are the spinless fermionic 
annihilation and creation operators.
Let the set of single-particle spatial functions constitute the molecular
 orbitals $\{\varphi(\textbf{r})\}^{M}_{k=1}$ and the set of spin orbitals $\{\chi(\textbf{x})\}_{p=1}^{2M}$ be defined with
$\chi_p=\varphi_i\sigma_i$ and the set of space-spin coordinates
$\textbf{x} = (\textbf{r}, \omega)$ where $\sigma_i$ is a spin function. The one-electron integral is defined as\cite{Daskin2}:
\begin{equation}
\begin{split}
h_{pq} &=\int{d\textbf{x} \chi_{p}^{*}(\textbf{x})\left(-\frac{1}{2}\triangledown^2-\sum_\alpha{\frac{Z_\alpha}{r_{\alpha \textbf{x}}}}\right)\chi_{q}(\textbf{x})}\\
&=\langle\varphi_p\mid H^{(1)}\mid\varphi_q\rangle\delta_{\sigma_p\sigma_q}
\end{split}
\end{equation}
and the two electron integral is:
\begin{equation}
\begin{split}
 h_{pqrs}&=\int{d\textbf{x}_1d\textbf{x}_2\frac{\chi_{p}^*(\textbf{x}_1)\chi_{q}^*(\textbf{x}_2)\chi_{s}(\textbf{x}_1)\chi_{r}(\textbf{x}_2)}{r_{12}}}\\
&=\langle\varphi_p\mid\langle\varphi_q\mid H^{(2)}\mid\varphi_r\rangle\mid\varphi_s\rangle\delta_{\sigma_p\sigma_q}\delta_{\sigma_r\sigma_s},
\end{split}
\end{equation}
where $r_{\alpha \textbf{x}}$ is the distance between the $\alpha^{th}$ nucleus and the electron, $r_{12}$ 
is the distance between electrons, $\triangledown^2$ 
is the Laplacian of the electron spatial coordinates, and $\chi_p(\textbf{x})$ is a selected single-particle basis:
 $\chi_p=\varphi_p\sigma_p$, $\chi_q=\varphi_q\sigma_q$, $\chi_r=\varphi_r\sigma_r$, and $\chi_s=\varphi_s\sigma_s$.

To describe the hydrogen molecule in minimal basis which is the minimum number of spatial functions required to describe the system, one spatial function is needed per atom denoted $\varphi
_{H1}$ and $\varphi
_{H2}$. 
The molecular spatial-orbitals are defined by symmetry: 
$\varphi_{g}=\varphi
_{H1}+\varphi_{H2}$ and $\varphi_{u}=\varphi
_{H1}-\varphi_{H2}$; which correspond to four spinorbitals:
$|\chi_1\rangle=|\varphi_g\rangle|\alpha\rangle, |\chi_2\rangle=|\varphi_g\rangle|\beta\rangle, |\chi_3\rangle=|\varphi_u\rangle|\alpha\rangle,$ and 
$|\chi_4\rangle=|\varphi_u\rangle|\beta\rangle$. The STO-3G basis is used to evaluate the spatial integrals of the Hamiltonian which is defined as
$\mathcal{H}=H^{(1)}+ H^{(2)}$, where since $h_{pqrs}=h_{pqsr}$, $H^{(1)}$ and $ H^{(2)}$ are simplified as\cite{Daskin2,Lanyon,Alan2}:\\

\begin{widetext}
\begin{equation}
\label{hamiltonianpart1}
H^{(1)}=h_{11}a_{1}^\dagger a_1+h_{22}a_{2}^\dagger a_2+h_{33}a_{3}^\dagger a_3+h_{44}a_{4}^\dagger a_4,
\end{equation}
and
\begin{equation}
\label{hamiltonianpart2}
\begin{split}
H^{(2)}&=h_{1221}a_{1}^\dagger a_{2}^\dagger a_{2} a_{1} +h_{3443}a_{3}^\dagger a_{4}^\dagger a_{4} a_{3} +h_{1441}a_{1}^\dagger a_{4}^\dagger a_{4} a_{1} + h_{2332}a_{2}^\dagger a_{3}^\dagger a_{3} a_{2}  +(h_{1331}-h_{1313})a_{1}^\dagger a_{3}^\dagger a_{3} a_{1}\\ &+(h_{2442}-h_{2424})a_{2}^\dagger a_{4}^\dagger a_{4} a_{2}+(h_{1423})(a_{1}^\dagger a_{4}^\dagger a_{2} a_{3} +a_{3}^\dagger a_{2}^\dagger a_{4} a_{1} )+(h_{1243})(a_{1}^\dagger a_{2}^\dagger a_{4} a_{3} +a_{3}^\dagger a_{4}^\dagger a_{2} a_{1} ).
\end{split}
\end{equation}
\end{widetext}

The spatial integral values evaluated for atomic distance $1.401 a.u.$,
 the Hamiltonian matrix found as a 16 by 16 matrix \cite{Daskin2}, so 4 qubits are required to implement the unitary propagator of this Hamiltonian which is found from $e^{-i\mathcal{H}t}$ by setting $t=1$. (see the note \footnote{For the matrix exponentiation, the MATLAB \textit{expm} function  which uses the Pade approximation
with scaling and squaring is used\cite{Daskin2}}).

The accuracy of the circuit design for the unitary propagator  also determines the accuracy of the simulation. 
The generation of quantum circuits by using matrix decomposition techniques or global optimization methods\cite{Daskin} (as done for water and hydrogen 
molecules in ref.\cite{Daskin2}) requires  searching a huge complex space and  simulation of the unitary matrices of quantum systems on classical computers. For large matrices, this 
hinders the efficiency, and hence, the accuracy of the circuits.
Since  the angles for the rotation gates in our circuits are determined from the matrix elements directly (for instance in the first design, Fig.\ref{circuit1full}), we only take the  
\textit{arcosine} of the values, and generating these angles requires only a few computations;   the  accuracy and the efficiency of the circuits are always high. This  helps to get very 
accurate circuit designs for the simulation of quantum systems.
For instance, for the 16 by 16 unitary propagator of hydrogen molecule given in ref.\cite{Daskin2}, nine qubits are required in the circuit scheme given in Fig.\ref{circuit1full}. Since the unitary propagator is highly sparse and  has only 19 nonzero elements,  most of the uniformly controlled gates in the circuit will be identity except 19 of them. Hence, in Appendix\ref{H2detail} we have shown how to reduce the number of qubits to 6 qubits, Fig.\ref{circuitH2}. We give the rotation values for the gates in Table \ref{table1}.  Therefore, since our circuit designs have fixed designs, using different basis sets or parameters to compute the Hamiltonian will not change the circuit design and the accuracy of it.  

In summary, we present general  programmable quantum circuits which can simulate any 
given $2^n$ by $2^n$ real matrix.  Because of the structure of the circuits, 
they can be used to fabricate specific or general purpose quantum chips and processors.
  Since the circuit designs  are highly dependent on the matrix elements;
for the application specific circuits aimed to implement particular type of systems,  any level of sparsity in the system may reduce the number of gates significantly. 
In addition, we show that the generation of circuits with the complexity less than the lower bound is possible by making $m\le2^{c-2}$ and increasing $\Phi$ in the given complexity. 
\section{Acknowledgments}
This work is supported by the NSF Centers for Chemical Innovation: Quantum Information  for Quantum Chemistry, CHE-1037992.

\appendix{
\section{The explicit illustration of the steps}
\label{ap}
Here,  we detail the implementation of the input modification, the formation ($V_f$), and the combination ($V_c$) steps. A sketch of the matrix format of the operations can be found in Eq.(\ref{eq:onequbit}) - for the one-qubit case in the first circuit design -  and Eq.(\ref{eq:twoqubit}) and Eq.(\ref{eq:twoqubit2}) - for the two-qubit case in the second circuit design; here blanks denote zeros and dots denote matrix parts of no interest for the final operation. 

\subsection{First circuit design}
\label{ap1}
Starting with an arbitrary input, $|\psi\rangle=(\alpha_0,\alpha_1)^T$, and the following arbitrary unitary matrix:
\begin{equation}
U = \left(
\begin{matrix}
u_{00}&u_{01}\\
u_{10}&u_{11}
\end{matrix}
\right)
\end{equation} 
the first method requires $2n+1=3$ qubits for the simulation with the input:
\begin{equation}
\label{ap:initial}
 |\psi_{initial}\rangle=|0\rangle\otimes|0\rangle\otimes|\psi\rangle=
\left(\begin{matrix}
\alpha_0\\
\alpha_1\\
0\\0\\0\\0\\0\\0
\end{matrix}\right).
\end{equation}

The followings represent the formation matrix, $V_f$, the matrix after the combination step, $V$ and the modified input, $|\tilde{\psi}\rangle$:
\begin{widetext}
\begin{equation}
V_f = \left(
\begin{matrix}
u_{00}&\cdot&&&&&&\\
\cdot&\cdot&&&&&&\\
&&u_{01}&\cdot&&&&\\
&&\cdot&\cdot&&&&\\
&&&&u_{10}&\cdot&&\\
&&&&\cdot&\cdot&&\\
&&&&&&u_{11}&\cdot\\
&&&&&&\cdot&\cdot
\end{matrix}
\right)
,\quad
V = \frac{1}{\sqrt{2}}\left(
\begin{matrix}
u_{00}&\cdot&u_{01}&\cdot&&&&\\
\cdot&\cdot&\cdot&\cdot&&&&\\
\cdot&\cdot&\cdot&\cdot&&&&\\
\cdot&\cdot&\cdot&\cdot&&&&\\
&&&&u_{10}&\cdot&u_{11}&\cdot\\
&&&&\cdot&\cdot&\cdot&\cdot\\
&&&&\cdot&\cdot&\cdot&\cdot\\
&&&&\cdot&\cdot&\cdot&\cdot
\end{matrix}
\right)
,\quad
|\tilde{\psi}\rangle = \frac{1}{\sqrt{2}}\left(
\begin{matrix}
\alpha_{0}\\
0\\
\alpha_{1}\\
0\\
\alpha_{0}\\
0\\
\alpha_{1}\\
0
\end{matrix}
\right)
\label{eq:onequbit}
\end{equation}
\end{widetext}

For illustration purposes, below we also present full forms of some of the operators and the output vector for the same case:

The full form of the resulting matrix from the formation step is as follows:
\begin{widetext}
\begin{equation}
V_f = \left(
\begin{array}{cccccccc}
 u_{00} & \sqrt{1-u_{00}^2} & 0 & 0 & 0 & 0 & 0 & 0 \\
 -\sqrt{1-u_{00}^2} & u_{00} & 0 & 0 & 0 & 0 & 0 & 0 \\
 0 & 0 & u_{01} & \sqrt{1-u_{01}^2} & 0 & 0 & 0 & 0 \\
 0 & 0 & -\sqrt{1-u_{01}^2} & u_{01} & 0 & 0 & 0 & 0 \\
 0 & 0 & 0 & 0 & u_{10} & \sqrt{1-u_{10}^2} & 0 & 0 \\
 0 & 0 & 0 & 0 & -\sqrt{1-u_{10}^2} & u_{10} & 0 & 0 \\
 0 & 0 & 0 & 0 & 0 & 0 & u_{11} & \sqrt{1-u_{11}^2} \\
 0 & 0 & 0 & 0 & 0 & 0 & -\sqrt{1-u_{11}^2} & u_{11}
\end{array}
\right)
\end{equation}
\end{widetext}

The combination matrix $V_C$ and the matrix for input modification $V_M$ are defined as:
\begin{widetext}
\begin{equation}
V_c = \left(
\begin{array}{cccccccc}
 \frac{1}{\sqrt{2}} & 0 & \frac{1}{\sqrt{2}} & 0 & 0 & 0 & 0 & 0 \\
 0 & \frac{1}{\sqrt{2}} & 0 & \frac{1}{\sqrt{2}} & 0 & 0 & 0 & 0 \\
 \frac{1}{\sqrt{2}} & 0 & -\frac{1}{\sqrt{2}} & 0 & 0 & 0 & 0 & 0 \\
 0 & \frac{1}{\sqrt{2}} & 0 & -\frac{1}{\sqrt{2}} & 0 & 0 & 0 & 0 \\
 0 & 0 & 0 & 0 & \frac{1}{\sqrt{2}} & 0 & \frac{1}{\sqrt{2}} & 0 \\
 0 & 0 & 0 & 0 & 0 & \frac{1}{\sqrt{2}} & 0 & \frac{1}{\sqrt{2}} \\
 0 & 0 & 0 & 0 & \frac{1}{\sqrt{2}} & 0 & -\frac{1}{\sqrt{2}} & 0 \\
 0 & 0 & 0 & 0 & 0 & \frac{1}{\sqrt{2}} & 0 & -\frac{1}{\sqrt{2}}
\end{array}
\right),\quad 
V_m = \left(
\begin{array}{cccccccc}
 \frac{1}{\sqrt{2}} & 0 & 0 & 0 & \frac{1}{\sqrt{2}} & 0 & 0 & 0 \\
 0 & 0 & \frac{1}{\sqrt{2}} & 0 & 0 & 0 & \frac{1}{\sqrt{2}} & 0 \\
 0 & \frac{1}{\sqrt{2}} & 0 & 0 & 0 & \frac{1}{\sqrt{2}} & 0 & 0 \\
 0 & 0 & 0 & \frac{1}{\sqrt{2}} & 0 & 0 & 0 & \frac{1}{\sqrt{2}} \\
 \frac{1}{\sqrt{2}} & 0 & 0 & 0 & -\frac{1}{\sqrt{2}} & 0 & 0 & 0 \\
 0 & 0 & \frac{1}{\sqrt{2}} & 0 & 0 & 0 & -\frac{1}{\sqrt{2}} & 0 \\
 0 & \frac{1}{\sqrt{2}} & 0 & 0 & 0 & -\frac{1}{\sqrt{2}} & 0 & 0 \\
 0 & 0 & 0 & \frac{1}{\sqrt{2}} & 0 & 0 & 0 & -\frac{1}{\sqrt{2}}
\end{array}
\right)
\end{equation}
\end{widetext}

For the initial input  $|\psi_{initial}\rangle$ as in Eq.(\ref{ap:initial}),
the final output state $|\psi_{final}\rangle$ becomes:

\begin{equation}
\begin{split}
|\psi_{final}\rangle&=V_cV_fV_m|\psi_{initial}\rangle\\
&=\frac{1}{2}\left(\begin{matrix}
 \alpha _{0} u_{00}+\alpha _{1} u_{01} \\
 - \alpha _{0} \sqrt{1-u_{00}^2}- \alpha _{1} \sqrt{1-u_{01}^2} \\
 \alpha _{0} u_{00}-\alpha _{1} u_{01} \\
 -\alpha _{0} \sqrt{1-u_{00}^2}+ \alpha _{1} \sqrt{1-u_{01}^2} \\
 \alpha _{0} u_{10}+\alpha _{1} u_{11}\\
 - \alpha _{0} \sqrt{1-u_{10}^2}- \alpha _{1} \sqrt{1-u_{11}^2} \\
 \alpha _{0} u_{10}-\alpha _{1} u_{11} \\
 - \alpha _{0} \sqrt{1-u_{10}^2}+ \alpha _{1} \sqrt{1-u_{11}^2}
\end{matrix}\right)
\end{split}
\end{equation}

Clearly the normalized states $|00-0\rangle$ and $|10-0\rangle$ simulate the original given system.

\subsection{Second circuit design}
\label{ap2}
For the same case, since the second circuit design initially works at least a pair of matrix elements, it will create the unitary at the initial step. There will be no need for the combination step. Hence, the output will be simulated on the states $|00\rangle$ and $|10\rangle$. For  two qubit system below, the simulation goes as follows:
\begin{equation}
U = \left(
\begin{matrix}
u_{00}&u_{01}&u_{02}&u_{03}\\
u_{10}&u_{11}&u_{12}&u_{13}\\
u_{20}&u_{21}&u_{22}&u_{23}\\
u_{30}&u_{31}&u_{32}&u_{33}\\
\end{matrix}
\right)
\end{equation}

In the formation step, if we use 4 by 4 blocks as shown in Fig.\ref{circuit2full}, there will be no need for the combination step since we will have already formed the rows of $U$ at the formation step. However,  if we  use 2 by 2 initial blocks, we need to use one rotation gate for each pair of the elements, then the combination step. Thus, at the end of the formation step,  we get the following matrix:
\begin{widetext}
\begin{equation}
 V_f = \left(
\begin{array}{cccccccccccccccc}
k_0u_{00}&k_0u_{01}\\
\cdot&\cdot\\
&&k_1u_{02}&k_1u_{03}\\
&&\cdot&\cdot\\
&&&&k_2u_{10}&k_2u_{11}\\
&&&&\cdot&\cdot\\
&&&&&&k_3u_{12}&k_3u_{13}\\
&&&&&&\cdot&\cdot\\
&&&&&&&&k_4u_{20}&k_4u_{21}\\
&&&&&&&&\cdot&\cdot\\
&&&&&&&&&&k_5u_{22}&k_5u_{23}\\
&&&&&&&&&&\cdot&\cdot\\
&&&&&&&&&&&&k_6u_{30}&k_6u_{31}\\
&&&&&&&&&&&&\cdot&\cdot\\
&&&&&&&&&&&&&&k_7u_{32}&k_7u_{33}\\
&&&&&&&&&&&&&&\cdot&\cdot
\end{array}
\right),
\label{eq:twoqubit}
\end{equation}
\end{widetext}
where $k_i$s are the normalization constants. After the sequential combination steps and the modification on the input, we get the following matrix and the modified input:

\begin{widetext}
\begin{equation}
V = \left(
\begin{array}{cccccccccccccccc}
u_{00}&u_{01}&u_{02}&u_{03}\\
\cdot&\cdot&\cdot&\cdot\\
\cdot&\cdot&\cdot&\cdot\\
\cdot&\cdot&\cdot&\cdot\\
&&&&u_{10}&u_{11}&u_{12}&u_{13}\\
&&&&\cdot&\cdot&\cdot&\cdot\\
&&&&\cdot&\cdot&\cdot&\cdot\\
&&&&\cdot&\cdot&\cdot&\cdot\\
&&&&&&&&u_{20}&u_{21}&u_{22}&u_{23}\\
&&&&&&&&\cdot&\cdot&\cdot&\cdot\\
&&&&&&&&\cdot&\cdot&\cdot&\cdot\\
&&&&&&&&\cdot&\cdot&\cdot&\cdot\\
&&&&&&&&&&&&u_{30}&u_{31}&u_{32}&u_{33}\\
&&&&&&&&&&&&\cdot&\cdot&\cdot&\cdot\\
&&&&&&&&&&&&\cdot&\cdot&\cdot&\cdot\\
&&&&&&&&&&&&\cdot&\cdot&\cdot&\cdot
\end{array}
\right)
,\quad
|\tilde{\psi}\rangle = 1/2\left(
\begin{matrix}
\alpha_{0}\\
\alpha_{1}\\
\alpha_{2}\\
\alpha_{3}\\
\alpha_{0}\\
\alpha_{1}\\
\alpha_{2}\\
\alpha_{3}\\
\alpha_{0}\\
\alpha_{1}\\
\alpha_{2}\\
\alpha_{3}\\
\alpha_{0}\\
\alpha_{1}\\
\alpha_{2}\\
\alpha_{3}
\end{matrix}
\right).
\label{eq:twoqubit2}
\end{equation}
\end{widetext}

The final state is equivalent to $|\psi_{final}\rangle=V|\tilde{\psi}\rangle$. In $|\psi_{final}\rangle$, the states $|0000\rangle, |0100\rangle, |1000\rangle,$ and $|1100\rangle$ are the respective states that simulate the original given unitary matrix. 
\section{Explicit Circuit for the unitary propagator of the Hydrogen Molecule}
\label{H2detail}
As mentioned, the  unitary matrix, $U_{H_2}$, for the hydrogen molecule has 19 nonzero elements, 15 of them 
located at the diagonal. Since the unitary is 16 by 16 we need 4 main and 5 ancilla qubits for the first circuit design given in
 Fig.\ref{circuit1full}. And the uniformly controlled rotation gates in the formation steps are the $R_y$ gates  followed by
 $R_z$ gates where we use identity for the zero elements. However, we can benefit from the sparsity of the matrix and
 reduce the number of ancilla to 2 qubits instead of 5: The non diagonal matrix elements are located at 
$(13,4), (4,13),(7,10),$ and $(10,7)$, where $(i,j)$ are the row and column indices.  We apply a permutation matrix, 
$P$, to reduce the bandwidth of the matrix. $PU_{H_2}$ takes non-diagonal elements $(13,4), (4,13),(7,10),$ and $(10,7)$ to $(5,4), (4,5),(7,8),$ and $(8,7)$ which creates another unitary, $\tilde{U}_{H_2}$. $\tilde{U}_{H_2}$ is a structured matrix where all the elements are located on the $(i,i)$, $(i,i+1)$, or $(i-1,i)$ positions. Hence, we can use  2 qubits for ancilla and 4 qubits for the main to create matrix $V$ having $4\times4$ block matrices  on its diagonal by using only one Hadamard gate in the combination step. In the formation step, the control qubits for $R_y$ gates and 
 $R_z$ gates are determined to form the couple of $(i,i)$ and $(i,i+1)$, or $(i-1,i)$ and $(i,i)$ elements on the first row of these 4 by 4 matrices. The angle values are determined from the polar representation of each element and given in Table \ref{table1}. The circuit for
 $\tilde{U}_{H_2}$ is shown in Fig.\ref{circuitH2} where $\tilde{R}$ represents a combination of an $R_y$ and an $R_z$ gates.  Please note that the circuit equivalences of the permutation matrices such as $P$ are the combinations of multi control CNOT gates where  which elements to be switched is determined by the control qubits. And the input should   be also permuted prior to the circuit. This can be done by simply switching the input for the qubits. At the end of this circuit,  since the leading rows of 4 by 4 matrices  simulate the unitary, we get the simulation result from the states $|\bold{0}\rangle,|\bold{4}\rangle,|\bold{8}\rangle,|\bold{12}\rangle,\dots,|\bold{60}\rangle$.
\begin{figure}[H]
  \centering
  \includegraphics[scale=0.3]{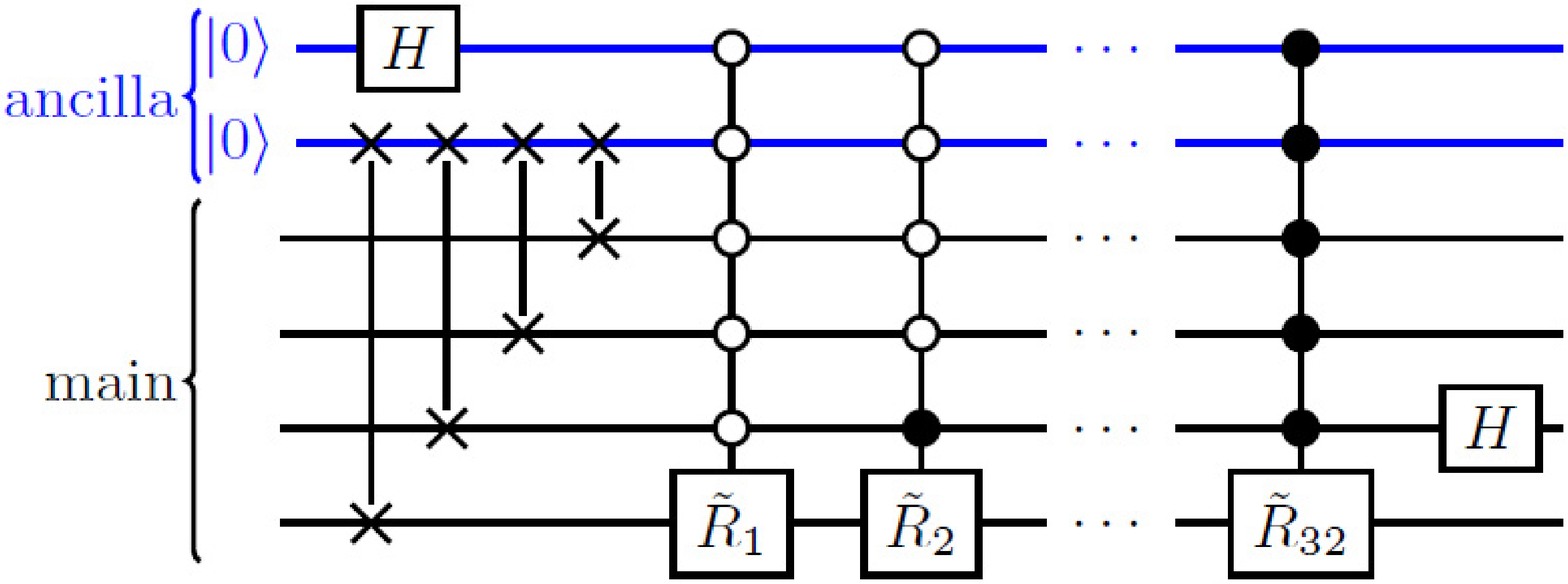}
  \caption{
  The circuit for the simulation of the hydrogen molecule. The angle values for the rotation gates are determined to create the elements of $\tilde{U}_{H_2}$: There is only 19 rotation gates, the rest is $X$ gates in order to get the right order for the elements after the combination . For diagonal elements of $\tilde{U}_{H_2}$, these rotations are only around $z$-axis. For nonzero-diagonal elements, rotation about z-axis followed by rotations about y-axis. The angles for these gates given in Table \ref{table1}.
  }
\label{circuitH2}
\end{figure}

\begin{table}[H]
\centering
\caption{Parameters for the Rotation Gates}
\label{table1}
\begin{tabular}{|p{0.5in}|c|p{0.5in}|p{0.4in}|}
\hline
State of  Control Qubits&Matrix Elements& Angle for $R_z$ & Angle for $R_y$\\ \hline
00000	&		0.9788-0.2049i		&	-0.4127	&		0	\\
00010	&		0.3987+0.9171i		&	2.3214	&		0	\\
00100	&		0.3987+0.9171i		&	2.3214	&		0	\\
00110	&		-0.2607+0.9517i		&	3.6763	&		0.3253\\
00111	&		0.1401-0.0817i		&	-1.0559	&		2.8158	\\
01000	&		0.1401-0.0817i		&	-1.0559	&		2.8158	\\
01001	&		-0.2607+0.9517i		&	3.6763	&		0.3253	\\
01011	&		0.9354+0.3535i		&	0.7226	&		0	\\
01101	&		0.3189+0.9478i		&	2.4925	&		0	\\
01110	&		0.4766+0.8604i		&	2.1299	&		0.3629\\
01111	&		-0.1577+0.0874i		&	5.271	&		2.779	\\
10000	&		-0.1577+0.0874i		&	5.271	&		2.779	\\
10001	&		0.4766+0.8604i		&	2.1299	&		0.3629	\\
10011	&		0.3130+0.9498i		&	2.5049	&		0	\\
10101	&		0.3189+0.9478i		&	2.4925	&		0	\\
10111	&		0.3130+0.9498i		&	2.5049	&		0	\\
11001	&		0.9569+0.2410i		&	0.4934	&		0	\\
11011	&		0.8889+0.4582i		&	0.9519	&		0	\\
11101	&		0.8889+0.4582i		&	0.9519	&		0	\\
11111	&		1		&	0	&		0	\\
\hline
\end{tabular}
\end{table}


%

\end{document}